\documentstyle[prl,twocolumn,aps]{revtex} 
\begin{document}
\draft
\title{Stability ordering of cycle expansions}
\author{C. P. Dettmann and G. P. Morriss }
\address{School of Physics, University of New South Wales, Sydney 2052,
Australia\\
c.dettmann@unsw.edu.au}
\date{\today}
\maketitle
\begin{abstract}
We propose that cycle expansions be ordered with respect to stability
rather than orbit length for many chaotic systems, particularly those
exhibiting crises.  This is illustrated with the strong field Lorentz
gas, where we obtain significant improvements over traditional
approaches.
\end{abstract}
Classical chaotic dynamical systems are inherently unpredictable, so
that even if the inital conditions are known with high accuracy,
predictions of the state of the system may be made to only a short time of
order $\lambda_{1}^{-1}$ in the future, where $\lambda_{1}$ is the
largest Lyapunov exponent.  Despite this unpredictability, much can be
said about the average behaviour of the system in many cases using only
a small number of unstable periodic orbits or ``cycles''.  This can be
achieved by using cycle expansions~\cite{C,AAC} of Ruelle's dynamical zeta
function~\cite{R,R2}.

Cycle expansions have proved very useful in both classical and quantum
chaos, giving accurate estimates of the escape rate of open billiard
systems~\cite{CE,GR,VWR} and the energy levels of helium~\cite{WRT}
using a surprisingly
small number of classical cycles.  The main idea of this
approach is that a long generic trajectory may be approximated by
various periodic orbits at different times, and that longer periodic
orbits may often be ``shadowed'' by shorter ``fundamental'' cycles,
closely following the shorter cycles along different sections of its
length.  Thus
averages are calculated using fundamental cycles, with small corrections
due to longer cycles.  Periodic orbits which are exact repetitions of
smaller cycles are explicitly summed, so that all expressions are written
in terms of the remaining ``prime'' cycles.
The expansions work best when the the symbolic dynamics is well
understood, and long periodic orbits are well shadowed by shorter ones.
In this paper we investigate a system in which neither of these
conditions holds,
the strong field Lorentz gas.  In spite of these difficulties,
reasonable results may be obtained by ordering the expansion in terms of
stability rather than the length of periodic orbits.  This approach
should be valid wherever cycle expansions can be applied, including
flows for which a natural topological ``length'' is difficult to
define.

The origin of cycle expansions is a well developed theory of trace
formulae and dynamical zeta
functions~\cite{R2,G}.  Here we will need only the expression for the
classical time average of some quantity $A$ in a closed system~\cite{AAC},
\begin{equation}
\langle A \rangle_t=\frac{\sum(-1)^k(A_1\tau_1+\cdots+A_k\tau_k)/
(\Lambda_1\cdots\Lambda_k)}
{\sum(-1)^k(\tau_1+\cdots+\tau_k)/(\Lambda_1\cdots\Lambda_k)}\;\;.
\label{aveq}
\end{equation}
The sum is over all distinct nonrepeating combinations of prime
cycles, $\tau$ is the period of a cycle, and $\Lambda$ is
magnitude of the expanding eigenvalue of the stability matrix, equal to
$\exp(\tau\sum\lambda_+)$, where $\sum\lambda_+$ is the sum of the
positive Lyapunov exponents. Note that for notational simplicity we have
written $\Lambda$ where other authors have $|\Lambda|$.
$A_i$ is the average of
some quantity over a particular orbit, and could be Lyapunov
exponent, current, or the number of collisions per unit time.  An
important check of the convergence of the expansion is given by
\begin{equation}
\zeta(0,0)^{-1}=1+\sum(-1)^k/(\Lambda_1\cdots\Lambda_k)
\label{normeq}
\end{equation}
which should be equal to zero.

The sums are usually truncated up to a particular order in the symbolic
dynamics: for the Lorentz gas, the number of collisions $N$.
At each order the terms are 
grouped into fundamental orbits that have a symbolic dynamics not
separable into smaller
cycles, and ``curvature corrections'' consisting of a composite orbit,
with symbolic dynamics $ab$ together with its components $a$ and $b$.
Often the composite orbit has similar averages to its components; then
the
combined term $-A_{ab}\tau_{ab}\Lambda^{-1}_{ab}+
(A_{a}\tau_{a}+A_{b}\tau_{b})\Lambda^{-1}_{a}\Lambda^{-1}_{b}$
is small.  If there are only a few fundamental orbits and this
cancellation occurs most of the time the cycle expansion converges
rapidly as a function of $N$.  However, this is often not the case,
as we see for the Lorentz gas (Tab.~\ref{orbits} below).

We suggest that the orbit stability $\Lambda$ may often be a more
convenient parameter with which to truncate the cycle expansions.
There are a number of reasons for this.  The cycles which make the greatest
contribution to the expansions are those with smallest $\Lambda$, which
are not necessarily those of smallest $N$.  The cycles of smallest
$\Lambda$ are also found most readily in numerical searches.
Alternative methods for enumerating periodic orbits by exhaustive
enumeration of symbol sequences (as in the zero field
Lorentz gas~\cite{CGS}) fail whenever the symbolic dynamics is not
understood completely and cycles exist which are not contained in the
chaotic set.  This is the case on one side of a chaotic crisis.  See
chapter~8 of Ref.~\cite{O}, where a number of examples are discussed,
including the windows of the logistic map, and the Lorenz attractor.
Another reason for truncating in $\Lambda$ rather than $N$ is that there
may be significant cycles (small $\Lambda$) at comparatively large $N$.
Continuing the expansion to include all cycles at large $N$ may be
prohibitive due to the large number of cycles, or that some of the
cycles may have very large $\Lambda$ and hence pose numerical problems.
These issues are illustrated by our study of the strong field Lorentz
gas.

We now make the above prescription more precise.  To retain the
exponential convergence of traditional cycle expansions, it is clear
that we must still make use of the close cancellation of curvature
correction terms.  This comes naturally if we include all terms for which the
product of $\Lambda$'s is less than a predetermined cutoff $\Lambda_{\rm
max}$.  Then, if the product of $\Lambda$'s match, the cancelling terms
will be either included or excluded together.  If the shadowing is poor,
cancellation would not occur whichever terms are included.  In any case,
it is clear that the expansions will not work well if cycles are
missing, whether the expansion is ordered with respect to $N$ or
$\Lambda$.  Our prescription is thus:

{\it Alternative truncation of cycle expansions: only those cycles
and combination of cycles which have the product of $\Lambda$'s less
than a cutoff $\Lambda_{\rm max}$ should be included in the cycle
expansion.  $\Lambda_{\rm max}$ is the largest value such that almost
all the cycles are known.}

For the remainder of this paper we show how this works for the strong
field Lorentz gas, obtaining significantly improved convergence.

\begin{figure}
\vspace*{3cm}
\includegraphics{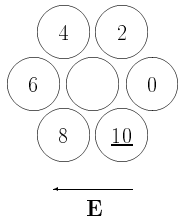}
\caption{\label{sd}The symbolic dynamics for the Lorentz gas at high
field, for which only nearest neighbour collisions occur.  We reduce the
symbol set still further by noting that $\underline{10}$ is equivalent
to $2$ by symmetry, and similarly $8$ is equivalent to $4$.}
\end{figure}

The two dimensional nonequilibrium Lorentz gas is a classical model of
current flow in a
conductor, and shares many of the properties of larger steady state
nonequilibrium systems, including a simple relationship between the
steady state current, and the sum of the Lyapunov exponents ${\bf J}\cdot
{\bf E}=\sum\lambda$~\cite{MH,LNRM}.  Here, $\bf E$ is the imposed electric
field.  A single particle of charge $e$ is scattered by a triangular
lattice of circular disks of radius $1$ and nearest neighbour spacing
$0.236$.  Between collisions it moves according to the equations
\begin{equation}
\dot{\bf x}={\bf v}\quad\dot{\bf v}=e{\bf E}/m-\alpha{\bf v}
\end{equation}
where $\alpha=e{\bf v}\cdot{\bf E}/(mv^2)$ is a Gaussian thermostatting
coefficient~\cite{EM} which keeps the kinetic energy constant, permitting
the system to reach a steady current ${\bf J}=\langle e{\bf v}\rangle_t$.
The symbolic dynamics is determined by the sequence of disks with which the
particle collides; see Fig.~\ref{sd}.

For small fields $\epsilon=|e{\bf E}/m|\ll 1$ the dynamics is known to
be ergodic and lead to a steady current~\cite{CELS}.  There have been
several previous investigations~\cite{CGS,V,MR} of the Lorentz gas
at zero field (equilibrium) and small fields
using both cycle expansions (above) and direct calculation of measures
based on periodic orbits.
They used $10^3$-$10^5$ cycles to evaluate the largest Lyapunov exponent,
the diffusion coefficient (for zero field), and the current (for nonzero
field).  In each case the expansions were evaluated to a particular
orbit length $N$, from 7 to 10.  Most of the results have uncertainties
of $5$-$10\%$, the main
exception being the evaluation of the Lyapunov exponent in
Ref.~\cite{CGS}, which has an uncertainty of about $0.2\%$.  This last
calculation was performed in the fundamental domain, making full use of
the symmetry of the lattice.  The convergence for the Lorentz gas
calculations is not as impressive as that in Refs.~\cite{CE,WRT} because the
grammar of the symbolic dynamics, that is, the rules determining which
symbol sequences occur, is not well understood.  Thus the amount to
which longer orbits are shadowed by shorter ones is limited.

At a field of $\epsilon=2.2$ the ergodic attracting measure is abruptly
replaced by an attractor with fractal support characterised by a very
restricted symbolic dynamics~\cite{DM}.  In particular (see
Fig.~\ref{sd}), the only symbols are $\{2,4,6,8,\underline{10}\}$, the
allowed pairs of symbols are $\{26,28,48,4\underline{10},62,64,68,
6\underline{10},82,84,\underline{10}4,\underline{10}6\}$, and the triples
$\{262,264,\underline{10}68,\underline{10}6\underline{10}\}$ are prohibited.
These rules determine that the particle moves in one direction along a
channel formed by the disks, as described in Ref.~\cite{DM}.  Making use
of the fact that the system is invariant under a reflection in the
$x$-axis, we note that the symbols $2$ and $\underline{10}$ are equivalent,
and likewise $4$ and $8$.  Thus the above permitted sequences become
$\{24,26,42,44,62,64\}$, with no restrictions on the triples.
It is easy to check that each symbol sequence of the restricted set
corresponds to exactly two equivalent sequences with the original
symbols.  Note that there are many other prohibited sequences of three or more
symbols, depending on the field, so this system suffers from pruning in a
similar fashion to the small field case.  The range of fields we consider
is $2.2$ to $2.4$, above which stable orbits appear, rendering the
attractor nonhyperbolic.

The symbolic dynamics of this system is so restricted that, at
$\epsilon=2.4$ there are only nine orbits with seven or fewer
collisions: see Tab.~\ref{orbits}.  There are a couple of points to note from
this table.  First, there is very little shadowing of orbits.  Five
of the cycles are fundamental in that they cannot be constructed
from smaller cycles.  In addition, the cycles that can be constructed
from smaller cycles do not have eigenvalues $\Lambda$ which are close to
the product of the constituent $\Lambda$'s, so curvature corrections
obtained with these orbits are not particularly small.  Secondly, the
most significant orbits (with small $\Lambda$) do not occur at small
$N$.  There are two $N=10$ orbits more significant than the
$N=2$ orbit, and 20 orbits with $\Lambda<100$, including one with
$N=34$.  Most of the longer orbits are partly cancelled in curvature
corrections, but it is clear from the numerical results (below) that they
are necessary to obtain optimal convergence.

It is clear that obtaining all periodic orbits with fewer than, say, 30
collisions is quite unfeasible, since $N=7$ is already approaching
the limit.  In any case, this is not in line with the basic philosophy
of describing the dynamics with as few cycles as possible.  Thus we
proceed to truncate the expansions using $\Lambda$ rather than $N$.

\begin{figure}
\vspace*{5cm}
\includegraphics{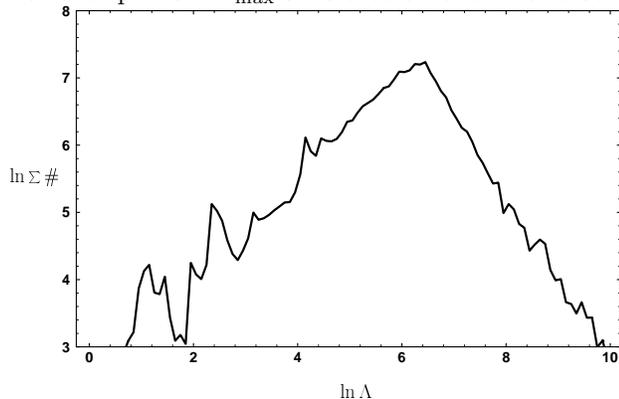}
\caption{\label{dist}The distribution of cycles found using our
numerical routine, using bins of equal size in terms of $\ln\Lambda$.
The maximum occurs at about $\Lambda=600$, suggesting this as the
optimal value of $\Lambda_{\rm max}$.}
\end{figure}

We numerically generated $10^5$ collisions for each of the field
values from $2.2$ to $2.4$ in steps of $0.001$, searching for periodic
orbits of up to $N=40$.  This generates about 170 cycles for each field
value, substantially fewer than Refs.~\cite{CGS,V,MR}, which considered a
much smaller number of field and disk spacing values.  Seven of the
orbits in Tab.~\ref{orbits} were found, those with the smallest
$\Lambda$.  To get a more precise estimate of the $\Lambda$
value at which cycles are first being missed, we plot the logarithm of
the total number of cycles found as a function of $\ln\Lambda$ in
Fig.~\ref{dist}.  The graph turns sharply down at about $\Lambda=600$,
indicating that the optimum $\Lambda_{\rm max}$ should be around this
value.

\begin{figure}
\vspace*{5cm}
\includegraphics{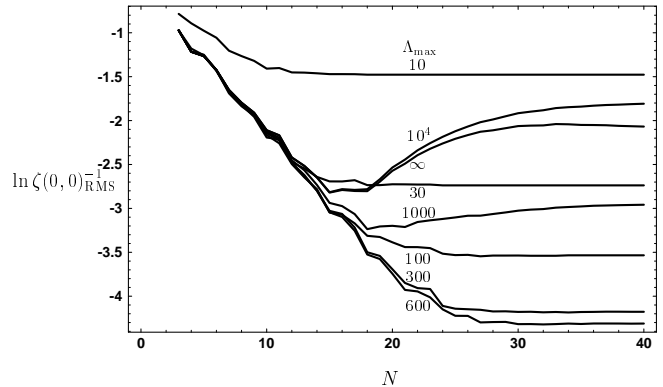}
\caption{\label{norm}Superior normalization of cycle expansions with
stability cutoff, using Eq.~(\protect\ref{normeq}).  All expansions with
$\Lambda_{\rm max}>10$ converge exponentially to $N=15$, after which the
expansion with no cutoff ($\Lambda_{\rm max}=\infty$) diverges slightly,
while the optimal cutoff ($\Lambda_{\rm max}=600$) continues to improve
up to $N=30$.}
\end{figure}

We computed $\zeta(0,0)^{-1}$ using Eq.~(\ref{normeq}) at each field
value summing terms up to different values of $N$ and $\Lambda_{\rm
max}$, and also evaluated the Lyapunov exponents $\lambda_1$,
$\lambda_2$ and the current $J$ using Eq.~(\ref{aveq}) under the same
conditions.  The RMS averaged $\zeta(0,0)^{-1}$ values are shown in
Fig.~\ref{norm}.  The cycle expansion values for $\lambda_1$,
$\lambda_2$ and $J$ were then compared with direct simulation
results at each field value, to give the RMS differences exhibited in
Tab.~\ref{hard}.

From Fig.~\ref{norm} we note that for each value of the cutoff, there is
an initial regime in which convergence is exponential with $N$, as usual
for cycle expansions in hyperbolic systems~\cite{AAC}.  At $\Lambda_{\rm
max}=600$ this holds to about $N=24$, while at $\Lambda_{\rm
max}=\infty$ (no cutoff), convergence is exponential only to $N=15$.
Subsequently the behaviour is
determined by the limitations of the data set.  If almost all of the
cycles are present (as for $\Lambda_{\rm max}<600$), the normalisation
improves slightly, and then remains constant, otherwise it gets worse.
A cutoff which is too high ($\Lambda_{\rm max}=10^4$)
can be more harmful than no cutoff, presumably because the unbalanced
corrections at $600<\Lambda<10^4$ are partly cancelled by the more
numerous
terms with opposite signs at larger $\Lambda$.

\begin{figure}
\vspace*{5cm}
\includegraphics{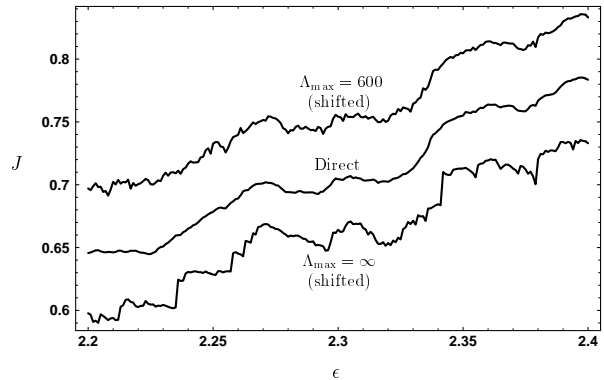}
\caption{\label{j}The current at high field (center).  Note the high
degree of structure, which is reproduced in the cycle expansions with
$\Lambda_{\rm max}=600$ (top) and to a lesser degree
$\Lambda_{\rm max}=\infty$ (bottom).
The cycle expansion results are shifted by $0.05$ for clarity.}
\end{figure}

The results for the Lyapunov exponents and the current
(Tab.~\ref{hard} and Fig.~\ref{j}) are more or less what would be
predicted from the
normalisation, except that the $\Lambda_{\rm max}=\infty$ results are
unexpectedly good, and comparable to $\Lambda_{\rm max}=100$.  Given
that $J\approx0.7$, the relative error in the cycle expansion expression
for $J$ at $\Lambda_{\rm max}=600$ is
about $0.3\%$, comparable to the best results in Ref.~\cite{CGS},
which required an order of magnitude more cycles.

\begin{figure}
\vspace*{5cm}
\includegraphics{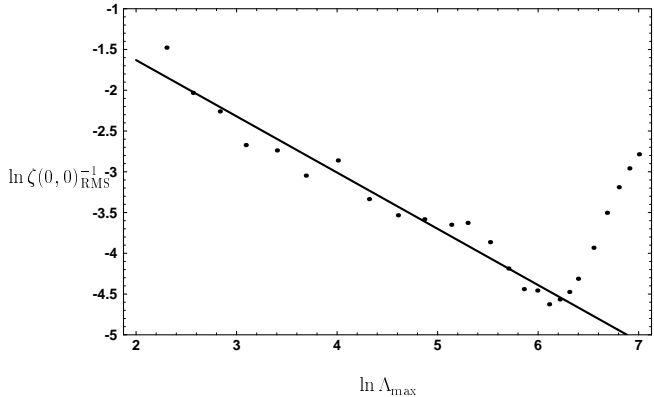}
\caption{\label{zl}The normalization at $N=40$ as a function of
$\ln\Lambda_{\rm max}$ showing exponential convergence up to
$\ln\Lambda_{\rm max}\approx6.3$.  The solid line has a gradient of
$-0.69$.}
\end{figure}

Finally, we use the results for $\zeta(0,0)^{-1}$ to estimate the rate
of convergence of the expansion when almost all the cycles are known,
that is, $\Lambda_{\rm max}<600$.  Note that because a hyperbolic system
has $\Lambda\sim e^{N}$, exponential convergence in $N$ is equivalent to
exponential convergence in $\ln\Lambda_{\rm max}$, that is a power law
in $\Lambda_{\rm max}$.  This is shown in
Fig.~\ref{zl}, where we find $\zeta(0,0)^{-1}_{\rm RMS}\sim\Lambda_{\rm
max}^{-0.69}$, or, $\zeta(0,0)^{-1}_{\rm RMS}\sim2^{-\ln\Lambda_{\rm max}}$.
The rate of convergence is thus comparable with that of
cycle expansions performed up to a given $N$.  The choice of which
approach to take depends on whether it is more convenient to enumerate
all orbits to a fixed $N$ or a fixed $\Lambda$, and the distribution of
periodic orbits for the given system.

It would be interesting to compare this method with the traditional
approach for other classical systems, and also quantum systems, where
formulae such as Eq.~(\ref{aveq}) are replaced by more involved
expressions, but still written in terms of classical periodic orbits.  
It could also be applied to flows for which no natural Poincare section
is known.  This work was supported by the Australian Research Council.

\begin{table}
\begin{tabular}{ccrc}
N&Symbolic Dynamics&\multicolumn{1}{c}{$\Lambda$}&$\tau$\\\hline
2&26&13.92~~~&1.459\\
3&264&2.557&2.706\\
4&2444&5.337&2.786\\
5&26264&60.50~~~&4.133\\
5&24426&175.6~~~~&3.061\\
6&244426&736.6~~~~&4.147\\
7&2442644&1.536&5.763\\
7&2444264&37.16~~~&5.471\\
7&2626264&907.9~~~~&5.594
\end{tabular}
\caption{\label{orbits}The shortest cycles at a field $\epsilon=2.4$.
Note the lack of shadowing, in that
$\Lambda_{244426}\gg\Lambda_{2444}\Lambda_{26}$.  Also, $\Lambda$ is not
strongly correlated with $N$, so that truncating a cycle expansion at
fixed $N$ tends to omit significant orbits with small $\Lambda$.  }
\end{table}

\begin{table}
\begin{tabular}{rrcccc}
$\Lambda_{\rm max}$&$\#_{\rm av}$&$\zeta(0,0)^{-1}_{\rm RMS}$&
$\Delta\lambda_{1\;\rm RMS}$&$\Delta\lambda_{2\;\rm RMS}$&$\Delta J_{\rm
RMS}$\\\hline
10&3.7&0.2283&0.0559&0.0865&0.0249\\
30&10.2&0.0647&0.0648&0.0774&0.0136\\
100&26.8&0.0292&0.0226&0.0312&0.0061\\
300&62.9&0.0153&0.0095&0.0125&0.0029\\
600&104.0&0.0134&0.0074&0.0102&0.0025\\
1000&131.1&0.0519&0.0162&0.0205&0.0039\\
10000&157.6&0.1640&0.1814&0.1525&0.0256\\
$\infty$&169.2&0.1264&0.0190&0.0331&0.0083
\end{tabular}
\caption{\label{hard}Cycle expansion results, showing the average number
of periodic orbits used, $\#_{\rm av}$, and the RMS differences of the
other
quantities from the direct simulation results.  All contributions up to
$N=40$ are included.  The best convergence is obtained with
$\Lambda_{\rm max}=600$, a factor of $3$ better than no cut off
($\Lambda_{\rm max}=\infty$).}
\end{table}

\end{document}